\documentstyle[prb,twocolumn,aps]{revtex}
\input{epsf}
\begin{document}
\draft
\twocolumn[\hsize\textwidth\columnwidth\hsize\csname @twocolumnfalse\endcsname
\title{A repulsive trap for two electrons in a magnetic field}

\author{A. D. Chepelianskii$^{(a)}$ and D. L. Shepelyansky$^{(b)}$}

\address {$^{(a)}$ Lyc\'ee Pierre de Fermat, Parvis des Jacobins, 31068 
Toulouse Cedex 7, France}
\address {$^{(b)}$ Laboratoire de Physique Quantique, UMR 5626 du CNRS, 
Universit\'e Paul Sabatier, 31062 Toulouse Cedex 4, France}

\date{November 1, 2000}

\maketitle

\begin{abstract}
We study numerically and analytically the dynamics of two classical electrons
with Coulomb interaction in a two dimensional antidot superlattice potential 
in the presence of crossed electric and magnetic fields. It is found that near
one antidot the electron pair can be trapped for a long time and the escape 
rate from such a trap is proportional to the square of a weak electric field.
This is qualitatively different from the case of noninteracting electrons 
which are trapped forever by the antidot. 
For the pair propagation in the antidot superlattice we found a broad 
parameter regime for which the pair is stable and where
two repulsive electrons propagate together on an enormously large distance.
\end{abstract}
\vskip1pc]

\narrowtext
\section{Introduction}

Recent technological development allowed to create various types of
surface superlattices for two-dimensional (2D) electron gas in semiconductor
heterostructures and to investigate their transport properties in
presence of magnetic field. Experiments with antidot lattices were carried out 
by different experimental groups (see e.g. \cite{ensslin,weiss,kvon,kvon1}) and 
the contribution of classical periodic orbits in the resistivity peaks
at certain values of magnetic field was clearly identified.
In fact the size of antidots and the distance between them are relatively 
large, and analysis of classical trajectory dynamics can be successfully 
applied to understand a number of unusual transport properties in such 
antidot arrays \cite{geisel1,geisel2}. Due to nonlinearity of motion in 
the vicinity of antidot potential the classical dynamics can be 
chaotic that leads to diffusive spreading of trajectories even for perfectly 
periodic lattices \cite{geisel1,geisel2}. If the distance between antidots
is large or comparable to the cyclotron radius of an electron in a magnetic
field perpendicular to the lattice, then one should understand first,
the properties of electron dynamics near one antidot. In the absence 
of electric field, the dynamics is integrable for an antidot of circular
shape due to angular momentum conservation and an electron always 
regularly rotates around the antidot. 
An electric field applied in the 2D plane of the superlattice
breaks the cylindrical symmetry and can lead to the electron escape from
the antidot to infinity. The problem of electron dynamics in crossed 
electric ${\em E}$ and magnetic ${\em B}$ fields near a circular elastic 
disk (antidot) was studied
in \cite{berglund}. It was shown that the dynamics can be described 
by a simple area-preserving map which depends only on one dimensionless 
parameter $\epsilon = (2 \pi m/a e)({\em E/B^2})$, where $m, e$ are 
electron mass and charge, and $a$ is the disk radius. For small 
$\epsilon < \epsilon_c$ the electron dynamics in the phase space of 
angular momentum $l$ and conjugated angle $\phi$ is bounded by the invariant
Kolmogorv-Arnold-Moser (KAM) curves so that electron always remains near
the disk. On the contrary for $\epsilon > \epsilon_c$ the KAM curves are
destroyed, global chaos sets in and the electron escapes to infinity
after few collisions with the disk.

Until now the classical dynamics in antidot lattices was 
studied only for noninteracting electrons \cite{geisel1,geisel2,berglund}.
In this paper for the first time we analyze the effect of Coulomb interaction
between classical electrons in the vicinity of an antidot. We show that 
for sufficiently strong interaction between electrons their dynamics 
becomes chaotic. Due to that one or two electrons can escape from the antidot 
even in an arbitrary weak applied electric field $E$ that corresponds 
to $\epsilon_c \rightarrow 0$ contrary to $\epsilon_c > 0$ in absence 
of interaction. We determine the dependence of average escape rate
$\Gamma$ on $\epsilon$ showing that in the limit of small electric field
$\Gamma \propto \epsilon^2$. After that we also discuss the two electron
propagation in the antidot superlattice.

The paper is organized as follows.
In the next Section we briefly discuss the one electron dynamics near the 
antidot in crossed magnetic and electric fields. 
In Section III the dynamics of two interacting electrons is analyzed 
in detail. The electron motion in the antidot superlattice is considered 
in Section IV.
In the last Section we summarize the obtained results.

\section{One electron dynamics}

The dynamics of an electron in crossed electric and magnetic fields
in two dimensions with one antidot is described by the Hamiltonian:
\begin{equation}
\label{ham}
H_0 = ({\mathbf p} - e {\mathbf A})^2 / 2 m + U (x,y) - e {\mathbf {E r}}
\end{equation}
where ${\mathbf A}$ is the vector potential
 and $U(x,y)$ describes the antidot potential
which depends only on the radius $r = \sqrt{x^2+y^2}$ with $ U = 0 $ for 
$ r \ge a $.
For convenience, following \cite{geisel1}, we introduce the dimensionless 
variables $\tilde{x}=x/a, \tilde{y}=y/a, \tilde{t}=t/\tau_0, 
\tilde{H_0}=H_0/2 \epsilon_F, \tilde{U}=U/2 \epsilon_F, \tilde{B} = B / B_0, 
\tilde{E}=E/E_0$
where $\epsilon_F (v_F) $ is the Fermi energy (velocity),
$\tau_0 = (2 \epsilon_F/m a^2 )^{-1/2} = a / v_F $,
magnetic and electric fields are scaled by 
$B_0 = (m \epsilon_F)^{1/2} / e a$ and $E_0 = 2 \epsilon_F /e a$ respectively.
In these units a magnetic field $B = B_0$ gives the cyclotron radius $R_c = a$
for electron with energy $\epsilon_F = m v^2_F /2 $.
We choose the Landau gauge ${\mathbf A} = (-By,0,0)$.
Then omitting the tildes the Hamiltonian equations of motions reads:
\begin{equation}
\label{equ}
\begin{array}{c}
d x/dt= v_x, \;\; d v_x/dt =  B v_y - d U/dx - E_x, \\
d y/dt= v_y, \;\; d v_y/dt = -B v_x - d U/dy - E_y 
\end{array}
\end{equation}
where $v_x = p_x + y B, v_y = p_y $ .
To model the antidot we chose the potential: 
\begin{equation}
\label{dotpot}
U (x,y) = U_0 (1 - r)^6 \;\;.
\end{equation}
Usually we choose $U_0$ to be much larger than the electron energy $H_0$
so that this potential becomes very similar to an absolutely rigid 
disk with effective radius $a_{eff}$ about 15\% smaller than $a$.

Far from the antidot the equations of motion are exactly solvable
and give electron rotation over a circle of cyclotron radius $R_c = v/\omega_c$
with cyclotron frequency $\omega_c = B$. In addition this circle 
moves with the drift velocity $v_d = E / B$ in the direction perpendicular
to electric and magnetic fields. As it was found by Berglund {\it et al.}
\cite{berglund}, near the antidot, the dynamics 
strongly depends on the dimensionless parameter 
$\epsilon = v_d 2 \pi / \omega_c = 2 \pi E / B^2$.
For $\epsilon \ll 1$ the electron scatters on the antidot and 
escapes to the infinity after one collision. 
On the contrary, the situation with not very large $\epsilon$
is much richer \cite{berglund}. In this case electron 
can collide many times with the antidot and this process is described by 
a simple area-preserving map \cite{berglund}:  
\begin{equation}
\label{map}
\begin{array}{c}
\bar{\phi}  = \phi  + \pi - 2 \sin^{-1} \beta \\
\bar{\beta} = \beta - \epsilon \sin \bar{\phi} 
\end{array}
\end{equation}
where bars denote the new values of variables after collision,
$\phi$ is the scattering angle measured in respect to the direction
of drift velocity and $\beta$ is the scattering impact parameter 
divided by the antidot radius. In this way $\beta$ varies in the 
interval (-1, 1). We note that $\beta$ can be also considered as 
the orbital momentum $l$ of the electron divided by the maximal momentum
$l_{max} = a v$ at which electron still collides with the antidot.
The real dynamics is correctly described by the map if $R_c \gg 1$,
that corresponds to $v \gg B$. For $\epsilon \gg 1$ the variation of
$\beta$ is bounded by the invariant KAM curves and electron is trapped 
near the antidot. The last KAM curve is destroyed for 
$\epsilon > \epsilon_c \approx 0.45\;\;$ \cite{berglund} so that 
the orbits with initial $\beta \approx 0$ can escape from the antidot
to infinity. Of course, for $\epsilon > \epsilon_c$ some islands
with regular motion inside still remain, but they become very small 
as soon as $\epsilon$ becomes significantly lager than $\epsilon_c$.

To study the electron dynamics in (\ref{ham}) the Hamiltonian 
equations of motion are solved numerically by Runge-Kutta 
method of fourth order so that the electron energy is conserved with the
relative precision better than $10^{-6}$. The examples of the Poincar\'e
cross sections constructed at $x = 0$ and $v_x > 0$ for
trajectories trapped near the antidot is shown in Figs. 1,2.
In Fig. 1 $\epsilon \approx 0.16$ is rather small and almost all
phase space is filled by integrable KAM curves. For Fig. 2, 
the parameter $\epsilon \approx 0.42$ is close to $\epsilon_c$
and KAM curves become more deformed and the chaotic component becomes
visible. This case can be compared with Fig. 3 in \cite{berglund}
where the cross section for the map (\ref{map}) is given for a close
value of $\epsilon$.

\begin{figure}
\epsfxsize=8cm
\epsfysize=6cm
\epsffile{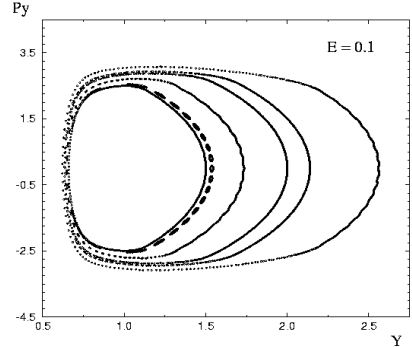}
\vglue 0.2cm
\caption{Poincar\'e cross section for the Hamiltonian (\ref{ham})
constructed at $x = 0, v_x > 0$ for $H_0 = 8.725, B = -2, E = 0.1$ so that
$\epsilon \approx 0.16$. The antidot determined by the potential (\ref{dotpot})
is located at (0, 0); $U_0 = 1000$.
} 
\label{fig1}
\end{figure}

\begin{figure}
\epsfxsize=8cm
\epsfysize=6cm
\epsffile{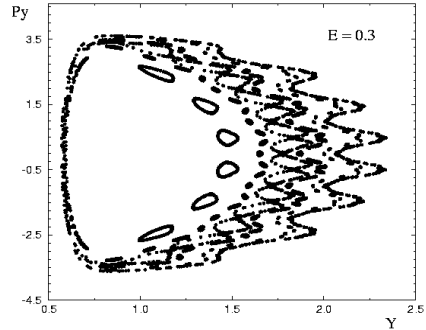}
\vglue 0.2cm
\caption{ Same as in Fig. 1 but for $E = 0.3$, $\epsilon \approx 0.42$.
} 
\label{fig2}
\end{figure}

\begin{figure}
\epsfxsize=8cm
\epsfysize=6cm
\epsffile{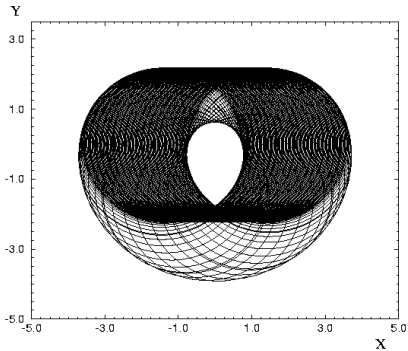}
\vglue 0.2cm
\caption{ Electron dynamics in $(x,y)$ plane near the antidot of Fig. 1 for
$H_0 = 9.7, B = -2.0, E = 0.1, \epsilon \approx 0.16$.
} 
\label{fig3}
\end{figure}

It is interesting to note that the map (\ref{map}) can be not valid if orbits
have $|\beta | \approx 1$ or $|\beta | > 1$. For example, if the
antidot is inside a large cyclotron circle then electron will make
many many rotations before this slowly drifting circle  will 
cross the antidot. This situation is not taken into account by the 
first equation in (\ref{map}). An example of the electron dynamics in 
this case is given in Fig. \ref{fig3}. Here $\epsilon \approx 0.16$
is small and the motion is still regular. We should stress that such
type of trajectories separates orbits which escape to infinity 
and those which collide with the antidot on each cyclotron period.
For the antidot superlattice with antidot spacing comparable with $R_c$
this type of orbits (see Fig. \ref{fig3}) 
is of special importance since these orbits can easily jump from 
one antidot to another leading to a global diffusion in the system.
We will discuss this situation in Section III.

\section{ Effects of Coulomb interaction on electron dynamics }

Let us now consider how the Coulomb interaction between two electrons 
affects their dynamics near the antidot. In this case the Hamiltonian
of the system reads 
\begin{equation}
\label{ham2}
H = H_0 ({\mathbf p_1}, {\mathbf r_1}) + H_0 ({\mathbf p_2}, {\mathbf r_2})
	+ e^2 / | {\mathbf r_1} - {\mathbf r_2} | \;\; .
\end{equation}
While in the free space the  Coulomb interaction repels the electrons 
and leads to their separation the situation 
is more complicated in the presence of magnetic field.
In the case without any antidot the total momentum 
of two electrons is conserved and as a result each electron rotates 
regularly on a cyclotron circle which in addition rotates around the
center of mass of the system. Without an external electric field 
($E = 0$) the center of mass is fixed and inert whereas in the 
presence of the field ($| E | > 0$) the center of mass drifts 
with the constant velocity $v_d = E / B$ but the average distance 
between electrons remains constant. However this electron pair can be 
trapped by the repulsive potential of the antidot so that the electrons 
will spend a long time colliding with this antidot. An example of 
electron dynamics in this case is shown in Fig. 4. It shows that an
electron can escape from the antidot even in the situation with
$\epsilon < \epsilon_c $ when without the interaction the electrons 
remain trapped near the antidot. In our numerical simulations we observed 
different cases where one or both of the electrons escape to infinity.
\begin{figure}
\epsfxsize=8cm
\epsfysize=6cm
\epsffile{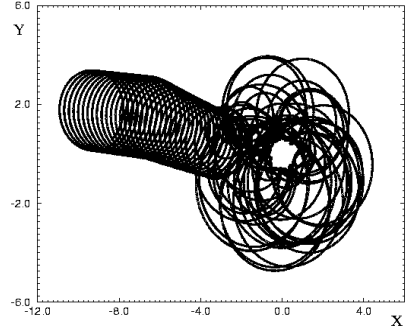}
\vglue 0.2cm
\epsfxsize=8cm
\epsfysize=6cm
\epsffile{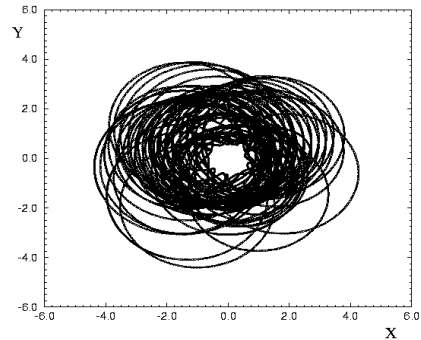}
\vglue 0.2cm
\caption{ Dynamics of two electrons in $(x,y)$ plane for 
$H = 15.65, B = -2.0, E = 0.15, \epsilon \approx 0.24$ and initial distance 
between electrons $| {\mathbf r_1} - {\mathbf r_2} | \approx 0.5 $.
After many cyclotron periods the first electron escapes from the 
antidot to infinity (upper figure) while the second remains trapped forever
(bottom figure).
} 
\label{fig4}
\end{figure}

To investigate how the escape rate depends on the strength of an
external electric field $E$ we studied the ensemble of 100 paths.
In each path the  positions and momentums of each electron are chosen 
randomly in the intervals $-4 \leq x, y \leq 4$, 
$ -2 \leq p_x, p_y \leq 2$ in such a way that the total energy is
$H \approx 15 \pm 0.5$. We remind that the antidot with potential 
(\ref{dotpot}) is placed at $(0,0)$ and $U_0 = 1000$. 
The escape rate $\Gamma$ is defined as $\Gamma = 1 / T$ where $T$
is the time after which the distance of one of the electrons from 
the antidot is greater than $R_{esc} \approx 5 R_c \approx 10$.
This distance is sufficiently large and as soon as it is reached 
an electron escapes to infinity and never returns back to the antidot.
The average value of $\Gamma$ is obtained by averaging over $100$ values
obtained for $100$ randomly chosen paths.

The dependence of the escape rate $\Gamma$ on the strength of the 
applied electric field $E$ is presented in Fig. 5. It definitely 
shows that the escape takes place even  at very weak electric fields 
with $\epsilon \ll \epsilon_c$ when without Coulomb interaction
electrons are forever trapped near the antidot.
According to the obtained numerical data (see Fig. 5) in the limit of
$\epsilon \rightarrow 0$ the escape rate is 
\begin{equation}
\label{gamma}
\Gamma / \omega_c \approx  \epsilon^2 \;\; .
\end{equation}

Our understanding of this dependence is based on the following argument.
Due to the Coulomb interaction between electrons their dynamics in 
the vicinity of the antidot becomes chaotic. Therefore the phase $\phi$
in the map (\ref{map}) changes randomly between electron collisions with 
the antidot and $\beta$ grows diffusively with the number of collisions
$n$ so that $(\Delta \beta)^2 \approx D n$ with $D = \epsilon^2 / 2$.
This diffusion results in the escape rate 
$\Gamma / \omega_c \sim D \sim \epsilon^2 $ being in agreement with the
numerical data in Fig. 5.
\begin{figure}
\epsfxsize=8cm
\epsfysize=6cm
\epsffile{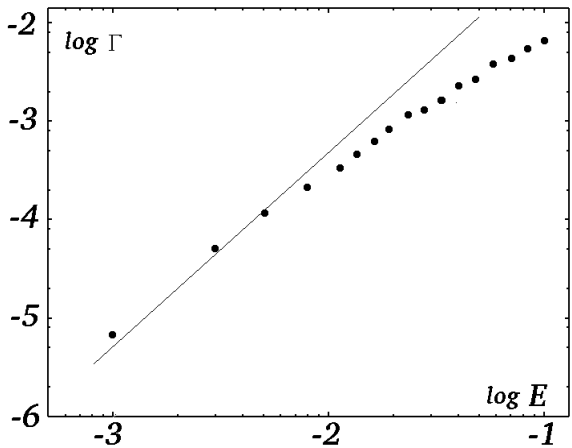}
\vglue 0.2cm
\caption{ Dependence of the escape rate $\Gamma $
on the electric field $E$ for electrons 
being initially in the vicinity of the antidot at 
$B = -2.0, H \approx 15.0 \pm 0.5, U_0 = 1000$. Here averaging 
is done over $100$ paths, $\omega_c =  2, \epsilon = 2 \pi E / B^2$,
points give the numerical results for $\Gamma  $ and the straight line
gives the dependence (\ref{gamma}). Logarithms are decimal.
} 
\label{fig5}
\end{figure}

\section{ Two electron propagation in antidot superlattice }

We also studied the electron dynamics on a square antidot superlattice, 
when the antidot potential is given by (\ref{dotpot}) and the distance 
between antidots $d > 2$. In this case our results for one electron dynamics 
are in qualitative agreement with the conclusions drawn in 
\cite{geisel1,geisel2}. As soon as the cyclotron radius $R_c$
becomes comparable with the antidot spacing $d$ the trajectories 
start to move diffusively on the whole lattice. Trapped orbits
near one antidot exist only for $2 R_c < d$  and $\epsilon < \epsilon_c$.
For $R_c > d / 2$ the cyclotron circle starts to drift 
in a way similar to that one shown in Fig. 3. After a time 
$t_d \sim d / v_d$ a collision with another antidot takes place 
that finally originates a sequence of irregular jumps between antidots.
The diffusion rate on the superlattice originated by this process
can be estimated as $D_{lat} \sim d^2/ t_d \sim d \; v_d$.
This diffusion is important in the limit $R_c \sim d \gg 1$.
However, we note that even at $E = 0$ at $R_c > d / 2$ there are chaotic 
orbits which diffuse over the whole lattice as it was discussed in detail
in \cite{geisel1,geisel2} and this diffusion is dominant for
$d \sim 1$.

It is interesting to understand how two electrons move in such a superlattice.
Intuitively, one would expect that the Coulomb repulsion will separate 
electrons and they will not propagate together. In fact we found that it 
is not necessarily the case and there are regimes where two electrons
propagate together. An example of such a case is shown in Fig. 6.
In this case the electron pair moves with an average drift velocity 
$v_d \approx E / B$ and the total displacement of the pair is about
hundred times larger than the distance between the two electrons
(Fig. 6). 
\begin{figure}
\epsfxsize=8cm
\epsfysize=6cm
\epsffile{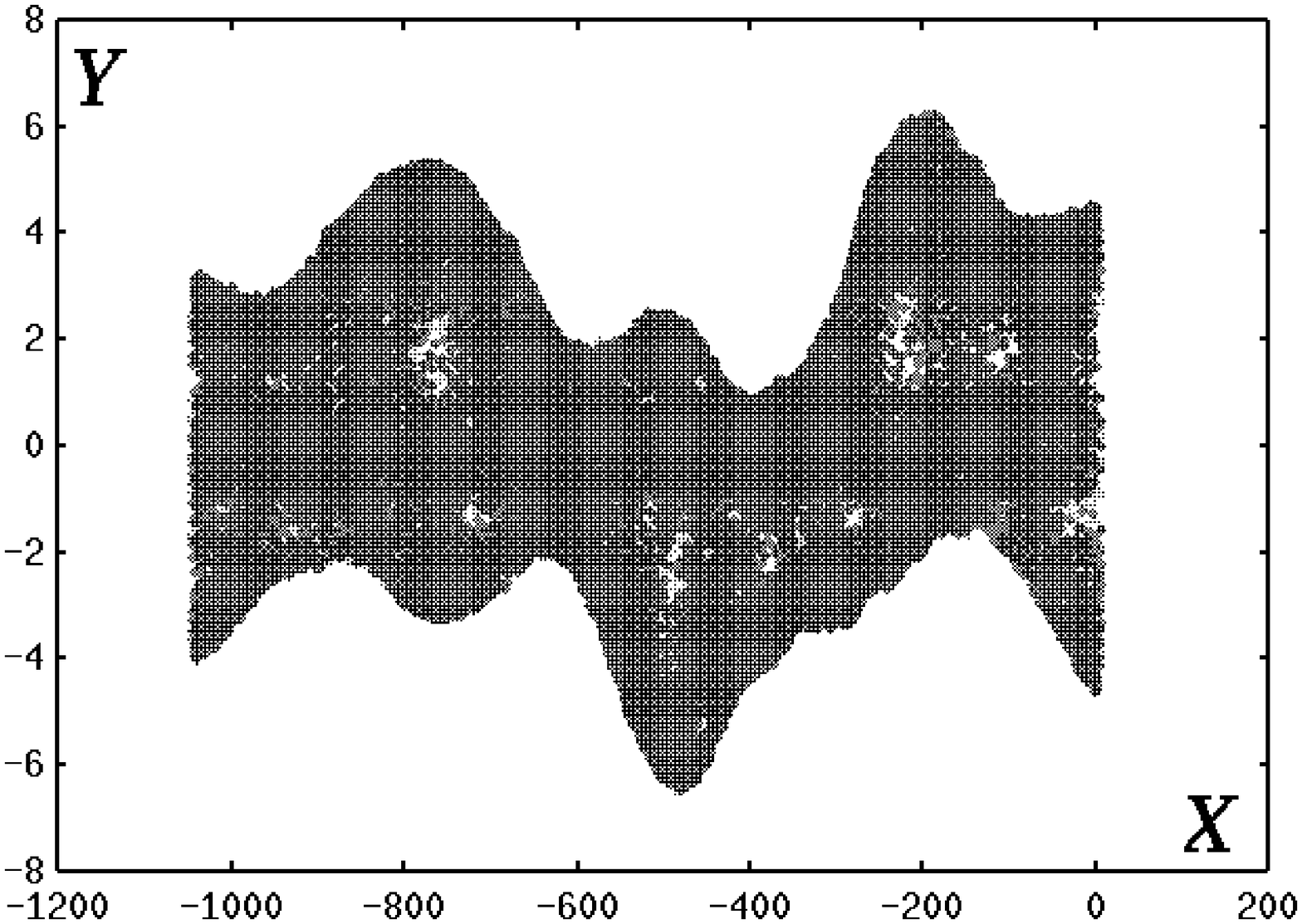}
\vglue 0.2cm
\epsfxsize=8cm
\epsfysize=6cm
\epsffile{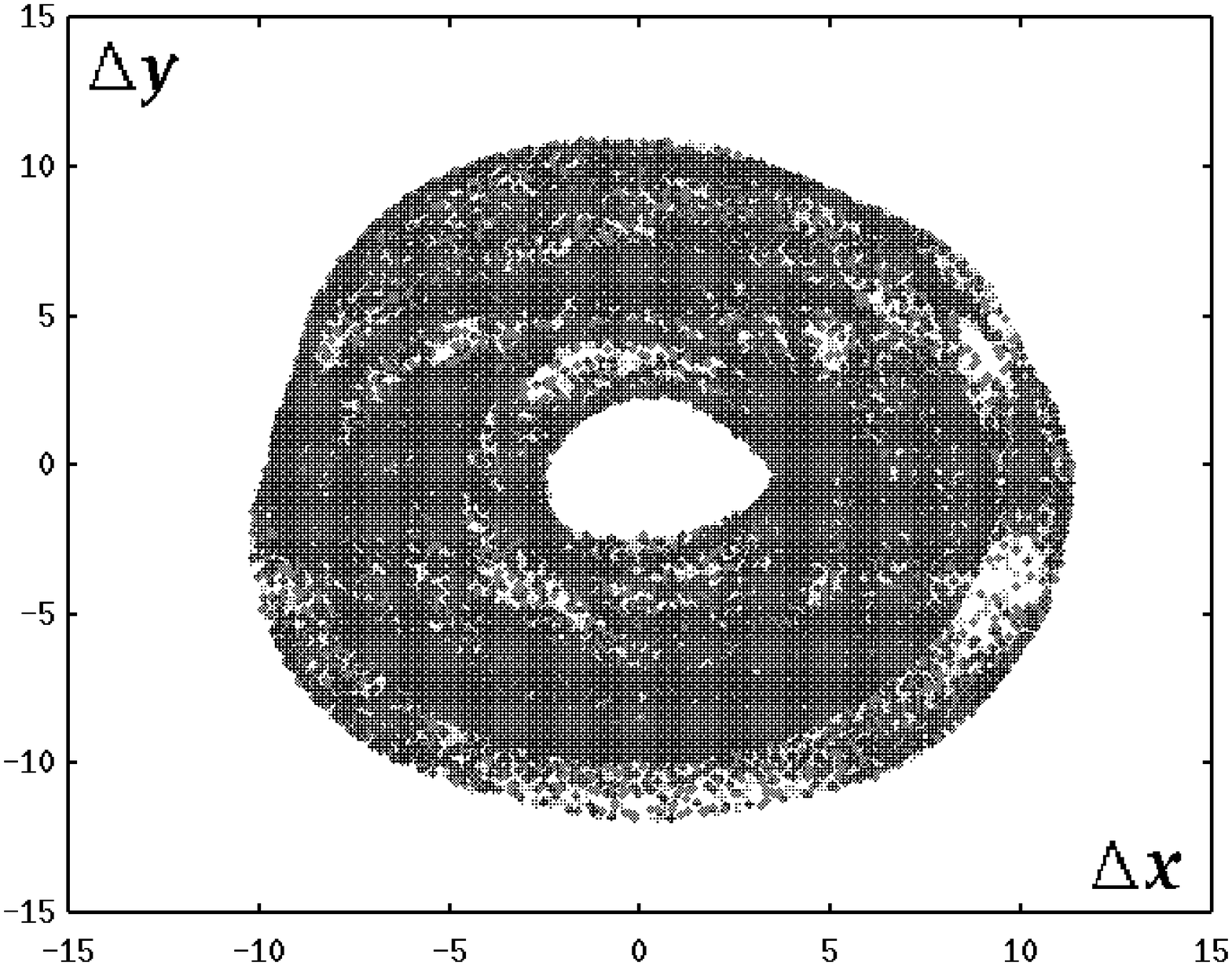}
\vglue 0.2cm
\caption{ Dynamics of two electrons in the $(x,y)$ plane of an antidot superlattice
for $H \approx 51, B = -2.0, E = 0.4, \epsilon \approx 0.63$ 
and initial distance between electrons 
$| {\mathbf r_1} - {\mathbf r_2} | \approx 4 $.
The antidots are placed on a square superlattice with period
4 and $U_0 = 1$ in (\ref{dotpot}). The upper figure shows the propagation
of the first electron in the plane $(x,y)$, while the bottom figure shows
the distance between the electrons $\Delta x = x_1 - x_2, \Delta y =y_1 -y_2$. 
} 
\label{fig6}
\end{figure}

The physical reason for appearance of such electron pairs is quite clear
in the absence of superlattice potential. In this case, as it was discussed
before at the beginning of Section III, electrons rotate one around another
and propagate together, their dynamics is integrable. Then
according to the KAM theorem a weak perturbation will not destroy such
pairs. Indeed, in the case of Fig. 6 the antidot potential is relatively 
weak ($U_0 = 1 \ll H / 2 \approx 25$) and the pair is not destroyed.
We checked numerically that provided $U_0 \sim H / 2$ the pair size starts
to grow diffusively due to random scattering on a strong antidot potential,
and eventually the pair is destroyed and electrons continue to propagate
separately. For $U_0 \gg H$ the separation happens after a few collisions
with antidots. 
On the contrary for $U_0 \ll H$ the life time of the 
classical pair becomes infinite in agreement with the KAM theorem.

\section { Conclusions }

In this paper we investigated the effects of Coulomb interaction between
two electrons on their classical dynamics in antidot superlattice
in crossed electric and magnetic fields. We found that for weak electric
field the electron pair can be trapped for a long time near an 
antidot even if eventually one or two electrons escape from the antidot.
The escape rate $\Gamma $ decreases proportionally to the square of electric
field. This behaviour is qualitatively different from the case of 
noninteracting electrons which are trapped forever near the antidot
in the limit of small electric field. 

The study of the electron pair dynamics in the antidot superlattice 
showed that the Coulomb repulsion can create stable pairs propagating 
on a large distance. In agreement with the KAM theorem such pairs 
are stable when the antidot potential strength is relatively weak 
compared to the electron energy. On the contrary, in the opposite 
limit the pairs become unstable and electrons are quickly separated 
from one another. On the basis of this phenomenon it is possible to
make a conjecture that in two dimensional heterostructures with high
mobility the impurity potential is relatively weak and such  electron 
KAM pairs will be stable and can be detected experimentally.
We note that in the recent experiments \cite{glattli} with 2D electron
gas the carriers of charge $2e$ have been detected.
It is possible that these carriers are  related to the KAM
pairs found in this paper. 

We thank G.Casati who attracted our interest to the results found in
\cite{berglund}.


\vskip -0.5cm
\begin{thebibliography}{99}
\bibitem{ensslin} K.~Ensslin and P.~M.~Petroff, Phys. Rev. B {\bf 41},
        12307 (1990).
\bibitem{weiss} D.~Weiss, M.~L.~Roukes, A.~Mesching, P.~Grambow,
        K. von Klitzing, G.~Weiman, Phys. Rev. Lett.
        {\bf 66}, 2790 (1991).
\bibitem{kvon} G.~M.~Gusev, V.~T.~Dolgopolov, Z.~D.~Kvon,
        A.~A.~Shashkin, V.~M.~Kudryashov, L.~V.~Litvin
        and Yu.~Nastaushev, Pis'ma Zh. Eksp. Teor. Fiz.
        {\bf 54}, 369 (1991).
\bibitem{kvon1} M.~V.~Budantsev, Z.~D.~Kvon, A.~G.~Pogosov, 
        G.~M.~Gusev, J.~C.~Portal, D.~K.~Maude, N.~T.~Moshegov
        and A.~I.~Toropov, Physica B {\bf 256-258}, 595 (1998).
\bibitem{geisel1} R.~Fleischmann, T.~Geisel, and R.~Ketzmerick,
        Phys. Rev. Lett. {\bf 68}, 1367 (1992).
\bibitem{geisel2} T.~Geisel, R.~Ketzmerick and O.~Schedletzky,  
        Phys. Rev. Lett. {\bf 69}, 1680 (1992).
\bibitem{berglund} N.~Berglund, A.~Hansen, E.~H.~Hauge, and J.~Piasecki, 
	 Phys. Rev. Lett. {\bf 77}, 2149 (1996).
\bibitem{glattli} D.~C.~Glattli, V.~Rodriguez, H.~Perrin,
        P.~Roche, Y.~Jin and B.~Etienne, Physica E
        {\bf 6}, 22 (2000).


\end{thebibliography}
\end{document}